\documentclass[aps,prb,twocolumn,superscriptaddress,showpacs,amsmath,amssymb,longbibliography]{revtex4-1}
\usepackage[english]{babel}
\usepackage{amsmath}
\usepackage{bm}
\usepackage{graphicx,bbm} 
\usepackage{times}
\usepackage{epsfig} 
\usepackage{soul}
\usepackage[utf8]{inputenc}
\usepackage[colorlinks,linkcolor=blue,citecolor=blue,urlcolor=blue]{hyperref}
\usepackage{color}
\usepackage{latexsym}
\usepackage{ulem}
\definecolor{nred}{RGB}{224,0,0}
\definecolor{nblue}{RGB}{28,130,185}
\definecolor{dgreen}{RGB}{78,138,21}
\definecolor{norange}{RGB}{230,120,20}

\newcommand{\tr}{\text{Tr}}

\newcommand{\cl}[1]{\hat{\mathcal{#1}}}
\begin{document} 
\title{From dissipationless to normal diffusion in easy-axis Heisenberg spin chain}
\author{P. Prelov\v{s}ek}
\affiliation{Jo\v zef Stefan Institute, SI-1000 Ljubljana, Slovenia}
\author{S. Nandy}
\affiliation{Jo\v zef Stefan Institute, SI-1000 Ljubljana, Slovenia}
\author{Z. Lenar\v{c}i\v{c}}
\affiliation{Jo\v zef Stefan Institute, SI-1000 Ljubljana, Slovenia}
\author{M. Mierzejewski}
\affiliation{Department of Theoretical Physics, Faculty of Fundamental Problems of Technology, Wroc\l aw University of Science and Technology, 50-370 Wroc\l aw, Poland}
\author{J. Herbrych}
\affiliation{Department of Theoretical Physics, Faculty of Fundamental Problems of Technology, Wroc\l aw University of Science and Technology, 50-370 Wroc\l aw, Poland}

\date{\today}
\begin{abstract}
The anomalous spin diffusion of the integrable easy-axis Heisenberg chain originates in the ballistic transport of symmetry sectors with nonzero magnetization. Ballistic transport is replaced by normal dissipative transport in all magnetization sectors upon introducing the integrability-breaking perturbations, including external driving. Such behavior implies that the diffusion constant obtained for the integrable model is relevant for the spread of spin excitations but not for the spin conductivity. We present numerical results for closed systems and driven open systems, indicating that the diffusion constant shows a discontinuous variation as the function of perturbation strength.
\end{abstract}
\maketitle

\section{Introduction} 
The transport in integrable quantum many-body models, the prominent example being the Heisenberg XXZ spin-$1/2$ chain, has attracted a lot of attention for a few decades and recently reached a high level of understanding via novel analytical and powerful numerical approaches \cite{bertini21}. However, there are still open fundamental questions, in particular, regarding the influence of integrability-breaking perturbations (IBP). While theoretical tools to deal with perturbations are well developed for the simpler but related problem of noninteracting fermions, perturbed integrable many-body systems still present a challenge \cite{bastianello21,mallayya19,lange18}.

The distinctive feature of integrable XXZ spin chain is the dissipationless ballistic transport at finite temperatures $T>0$ \cite{castella95,zotos96}. In the dynamical conductivity \mbox{$\sigma(\omega)=2\pi D\,\delta(\omega)+\sigma_{reg}(\omega)$} it is detected via finite Drude weight $D(T>0) >0$. While in general case the lower bound for $D$ can be related to the local conserved quantities (CQ) of integrable models \cite{zotos97}, for the most investigated case of easy-plane Heisenberg chain ($\Delta <1$) one has to invoke also quasi-local CQ \cite{prosen11,prosen13,ilievski16} to explain $D>0$. A more general description of integrable transport is recently developed framework of generalized hydrodynamics (GHD) \cite{bertini16,castro-alvaredo16,ilievski17,bulchandani18,denardis18,denardis19,gopalakrishnan19}. However, much less is understood about the role of IBP, which plausibly reduce the ballistic transport into a normal dissipative one \cite{castella96,zotos96,jung06,jung07,znidaric20}, attributed to the decay of CQ in the perturbed system \cite{mierzejewski15}. Transport in spin chains has been recently also realized and investigated experimentally in cold-atom systems \cite{hild14,jepsen20,wei22} 

A particular challenge is the easy-axis ($\Delta>1$) XXZ model where $D =0$ at zero magnetization $m=0$, whereas the nonequilibrium steady state of open, boundary driven systems \cite{znidaric11} and the Hamiltonian evolution of an inhomogeneous spin profile \cite{steinigeweg09,steinigeweg12,karrasch14,steinigeweg15} are consistent with a finite diffusion constant ${\cal D}_I >0$. It has become increasingly evident that such transport is anomalous. Namely, integrable models do not exhibit internal dissipation \cite{prelovsek04}. The dc conductivity $\sigma_0=\lim_{F \to 0} j_s/F$ is defined via the steady current, $j_s$,  induced by a steady  field, $F$, hence the conductivity is determined in the presence of nonzero driving. Such driving breaks integrability and restores normal transport (i.e., steady driving leads to a steady heating). More recently, the anomalous origin of spin diffusion within the integrable model has emerged also from GHD methodology, which explains diffusion constant ${\cal D}_I >0$ as the grandcanonical (GC) superposition of ballistic (i.e. dissipationless) spreading of quasiparticles belonging to $m \ne 0$ sectors \cite{medenjak17,ilievski18,gopalakrishnan19}. 

In this paper we investigate the effect of weak IBP of strength $g$ on the high-$T$ diffusion in the XXZ model, with the main conclusion that the dc diffusion constant ${\cal D}$, appearing at nonzero $g>0$ is unrelated to the dissipationless diffusion ${\cal D}_I$ in the integrable system, as previously hinted \cite{mierzejewski11,steinigeweg12} and recently also stated more explicitly \cite{denardis21a}. As a consequence, the generalized Einstein relation ${\cal D} = \sigma_0/ \chi_0$, which relates $\sigma_0$ and the spin susceptibility $\chi_0(T\to\infty)\sim 1/(4T)$ (taking $k_B=1$) is not valid for  integrable XXZ model, but restored in the presence of IBP. We establish this by studying different IBP and calculating dissipative/normal dc ${\cal D}(g)$ emerging from different methods: (a) the dynamical diffusion response in finite systems with $L \leq 32$ for different magnetization sectors $m \ge 0$, (b) the steady-state currents in open, boundary-driven systems allowing for much larger $L \leq 80$, (c) the time-dependent spin structure factor $S(q,t)$ at finite wavevectors $q=2\pi/L$, (d) the variation of modulated spin profile under the influence of finite driving field $F > 0$.

In the large-$L$ limit, results for ${\cal D}(g)$ of all methods are expected to converge (shown here for modest $g>0$). Most important, results indicate that ${\cal D}_0 = {\cal D}(g \to 0) \ll {\cal D}_I$, i.e., diffusion reveals a jump upon introduction of IBP \cite{denardis21}. On contrary, the Einstein relation (based on the linearity of system's response to weak IBP) is not valid for the integrable system and different protocols, e.g., the sequence of limits $L \to \infty$ and $\omega \to 0$ ($t \to \infty$) may yield different results.

\section{Model and dissipationless diffusion}
We study $S=1/2$ spin chain of length $L$ with anisotropy $\Delta >1$,
\begin{equation}
H= J \sum_{i} \left[ \frac{1}{2} \left( S^+_{i+1} S^-_i + \mathrm{H.c.} \right) + 
\Delta S^z_{i+1} S^z_i \right] + g H'\,.
\label{xxz}
\end{equation}
We consider different IBP that conserve the translational symmetry of the model. In the main text we focus on the staggered field $g=\delta h$, $H' = \sum_i (-1)^i S^z_i$, while results for staggered exchange $g=\delta J$, $ H' = (1/2) \sum_i (-1)^i (S^+_{i+1} S^-_i + \mathrm{H.c.}) $, and the next-nearest-neighbor (NNN) interaction $g=\Delta_2$, $H'= \sum_i S^z_{i+2} S^z_i$ are described in more detail in Appendix~\ref{app1} and~\ref{app2}. Note that, considered $H'$ conserve $S^z_{tot}= \sum_i S^z_i = m L$, where $|m| \leq 1/2$ is the magnetization of the system. We further use $J=1$ and focus on the uniform spin current, which is for unperturbed $H$ (also for $g=\delta h, \Delta_2$) given by $j_s=(J/2)\sum_{l}(i S^+_{l} S^-_{l+1} + \mathrm{H.c.})$.

Before presenting results for perturbed XXZ chain, we provide a simple interpretation of the dissipationless diffusion emerging in the unperturbed system at $\Delta>1$. We first notice that for latter system the Drude weight $D$ for $m \ll 1$ is dominated by the overlap of $j_s$ with the conserved energy current $j_E$ \cite{zotos97},
\begin{equation}
T D \gtrsim \frac{1}{2L} \frac{ \langle j_s j_E \rangle^2}{ \langle (j_E)^2 \rangle} = \frac{ \Delta^2 m^2} { 1+ 2 \Delta^2}\,.
\label{mazur}
\end{equation}
In an infinite system, this leads to ballistic component in each $m \ne 0$ sector. However, in a finite system, the spin current acquires a finite relaxation time $\tau_L$ due to scatterings at the system boundaries. Introducing an effective velocity for the ballistic spin propagation, $\bar v$, one gets the corresponding scattering rate, $1/\tau_L=\bar v/L$. Consequently, the $\delta(\omega)$-contributions are broaden to Lorentzians with the widths $\Gamma_L = 1/\tau_L$. It leads to a finite (diffusive-like) response, $\sigma_m (\omega \to 0) = 2 D(m)/\Gamma_L$, which is obviously only a finite-size effect in each sector $m \ne 0 $. Nevertheless, the grand-canonical (GC) averaging over $m$ effectively eliminates the size-dependence
\begin{equation}
{\cal D}_I = \frac{ 2 L \langle D(m) \rangle_{GC}}{\bar v \chi_0} \sim \frac{4 L}{\bar v} \langle m^2 \rangle_{GC}=\frac{1}{\bar v}\,,
\label{di}
\end{equation}
where we take into account $\Delta \gg 1$ in Eq.~(\ref{mazur}) and high-$T$ average $\langle m^2 \rangle_{GC} =1/(4L) $. The underlying assumption is that the effective $\bar v$ weakly depends on $m$. We note that analogous GC analysis is also the basis of the lower bound \cite{medenjak17} as well as the GHD result \cite{ilievski18,gopalakrishnan19} for ${\cal D}_I$.

This simple scenario introduces two crucial implications:
\begin{itemize}
\item Since the diffusive spin propagation originates from the ballistic transport in $m\ne0$ sectors, the diffusion is not related to any internal energy dissipation. It means that Eq.~(\ref{di}) is relevant for the propagation of spin excitations, but not for their dc response to a steady driving \cite{mierzejewski11} since the latter is immanently connected with the internal energy dissipation (c.f. the Joule heating). 
\item The IBP introduces an $L$-independent mean free path, $\lambda^*$,  which determines the scattering rate, $\tau^{-1}_\lambda=\bar v/\lambda^*$ and should be used instead of $L$ in the numerator in Eq.~(\ref{di}). Then, Eq.~(\ref{di}) becomes just a finite-size contribution that vanishes as $\lambda^* /L\to0$. In other words, even weak perturbation eliminates the contribution given by Eq.~(\ref{di}), leading to a discontinuous ${\cal D}$ in the thermodynamic limit.
\end{itemize}

\section{Dynamical diffusion in closed systems}

We calculate $T \gg 1$ dynamical conductivity $\tilde \sigma(\omega) = T \sigma(\omega)$ for systems with IBP with the linear-response (Kubo) relation
\begin{equation}
\tilde \sigma(\omega) = \frac{\pi}{L N_{st} }\sum_{n,m} |\langle n| j_s| m\rangle|^2 \delta
(\omega - \epsilon_m + \epsilon_n)\,,
\label{kubo}
\end{equation} 
where $|n\rangle,|m\rangle$ are $N_{st}$ many-body eigenstates with corresponding energies $\epsilon_ n,\epsilon_m$. In $L$-site chains with PBC and fixed (canonical) magnetization $m$ we evaluate $\tilde \sigma(\omega)$ using the microcanonical Lanczos method (MCLM) \cite{long03,prelovsek13}. The method is upgraded \cite{prelovsek21} by performing large number of Lanczos steps, i.e. $M_L \sim 5000$. This allows for resolution $\delta \omega \sim 10^{-3}$ for systems up to $L = 32$ with $N_{st} \lesssim 10^7$. Dealing with perturbed/normal systems we adopt the Einstein relation, ${\cal D}(\omega)= \sigma(\omega)/\chi_0 = 4 \tilde \sigma(\omega)$, as valid for $T \gg 1$ and $m \sim 0$.

\begin{figure}[tb]
\includegraphics[width=0.9\columnwidth]{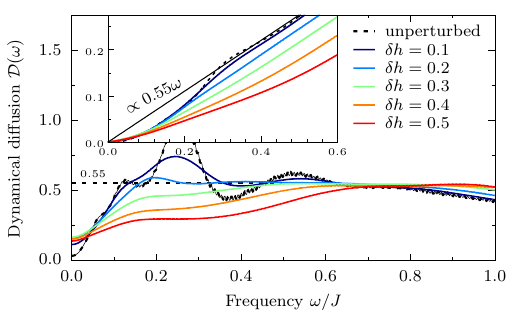}
\caption{ High-$T$ dynamical diffusion ${\cal D}(\omega)$ evaluated for XXZ chain at magnetization $m=0$ and fixed $\Delta = 1.5$, using MCLM on $L=32$ system and various staggered fields $\delta h$. The inset shows the integrated quantity $I(\omega) = \int_0^\omega d \omega' {\cal D}(\omega')$, with the extrapolation to limit $L \to \infty$ for the unperturbed system.}
\label{fig1}
\end{figure}

We first comment on the canonical $m=0$ results for ${\cal D}(\omega)$, as presented in Fig.~\ref{fig1}, for particular anisotropy $\Delta=1.5$ and calculated for maximum reachable $L=32$ for various staggered field $\delta h$. Analogous results for $\Delta_2>0$ and $\delta J>0$ are presented Appendices \ref{app1} and \ref{app2}, respectively. For comparison, we include the unperturbed $\delta h =0$ result obtained by the same protocol, with the inset showing the integrated spectra $I(\omega) = \int_0^\omega d \omega' {\cal D}(\omega')$. As already noticed \cite{prelovsek04}, the integrable case exhibits vanishing dc value ${\cal D} (\omega \to 0) \sim 0$, but as well large $\omega \sim 1/L$ oscillations, which can still be made compatible (after $L \to \infty$ extrapolation) with the result ${\cal D}_I \sim 0.55$ (see the inset), obtained also by other numerical approaches \cite{znidaric11,steinigeweg11,karrasch14,bertini21}, being also in the range of the GHD result \cite{gopalakrishnan19,denardis19}. 

On the other hand, results for finite $g =\delta h >0$ reveal dc value ${\cal D}(\omega\to0)={\cal D}_0 >0$, evidently smaller than the integrable one, ${\cal D}_0(g) \ll {\cal D}_I$. Such values are not finite-size effects, which can be deduced also from the integrated spectra $I(\omega)$, which are for $g > 0$ systematically reduced, i.e., $I(\omega) < {\cal D}_I \omega$ in a wide range $\omega <1$, requiring the reduction of ${\cal D}(\omega \sim 0)$. Some caution is appropriate for weakest perturbation, e.g., for $g \le 0.1$, where ${\cal D}(\omega)$ in Fig.~\ref{fig1} still exhibits visible finite-size oscillations $\omega \sim 1/L$. This indicates that reachable systems might be too small for considered $g$, i.e., $L < \lambda^*$ where $\lambda^*(g)$ represents an effective mean-free path. Similar are conclusions for $g=\Delta_2,\delta J$ perturbations shown in Appendices \ref{app1} and \ref{app2}, where $L$ dependence of ${\cal D}(\omega)$ spectra is presented.

\begin{figure}[tb]
\includegraphics[width=0.9\columnwidth]{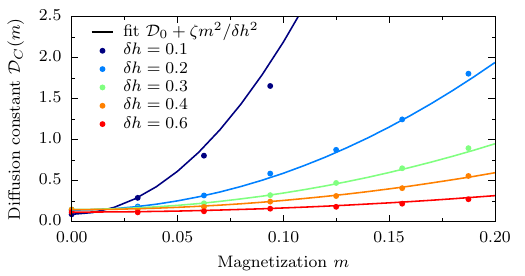}
\caption{Canonical dc diffusion constant ${\cal D}_C(m)$ for $\Delta = 1.5$ as calculated for different magnetizations $m$ with MCLM in a system of $L=32$ sites for different staggered fields $\delta h>0$. Results are fitted with parabolic dependence, Eq.~\eqref{dm} with fixed $\zeta =2.1$.}
\label{fig2}
\end{figure}

For given $L$, we employ MCLM to calculate canonical ${\cal D}_C(\omega,m)$ and extract dc ${\cal D}_C(\omega\to0,m)={\cal D}_C(m)$ corresponding to different $m$ sectors. Results presented in Fig.~\ref{fig2} show parabolic dependence on $m$, at least for $g \ll 1$. This can be explained as follows. Considered IBP do not conserve $j_E$ and lead to the normal (finite) dc energy/thermal conductivity $\kappa_0$, which scales (as expected from perturbation theory) as $\kappa_{0}\propto 1/g^2$~ \cite{jung06,jung07,znidaric20, mierzejewski21}. Taking into account the overlap of the spin current with $j_E$, Eq.~(\ref{mazur}), this leads to the dependence,
\begin{equation}
{\cal D}_C(m) = {\cal D}_0 + \zeta m^2/g^2. \label{dm}
\end{equation}
Results in Fig.~\ref{fig2} are well represented with the above dependence, where $\zeta$ appears universal for all $g >0$, plausibly still dependent on the type of perturbation, e.g., $\zeta = 2.1$ for $g=\delta h$, but $\zeta = 1.8, 9.0$ for $g=\Delta_2,\delta J$ cases, respectively (Appendices \ref{app1} and \ref{app2}). The ${\cal D}_0={\cal D}_C(m=0)$ represent weakly $g$ dependent canonical $m=0$ values, extracted from Fig.~\ref{fig1}, and apparently determined by a mechanism unrelated to $j_E$.

\begin{figure}[tb]
\includegraphics[width=0.9\columnwidth]{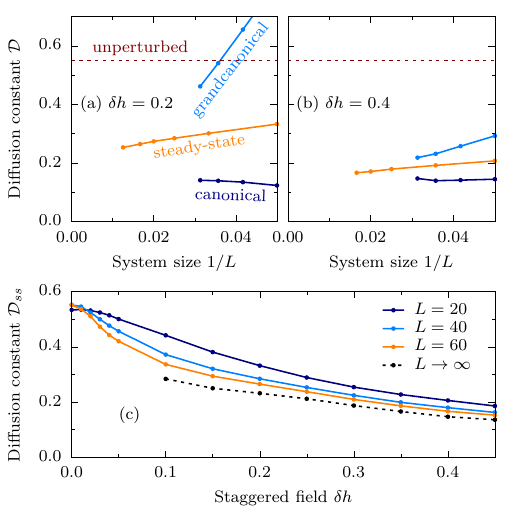}
\caption{Diffusion constant ${\cal D}$ vs. $1/L$ obtained for fixed $\Delta = 1.5$ from finite-$L$ results by three different approaches: canonical $m=0$ sector MCLM results ${\cal D}_0$, GC average over $m$ sectors ${\cal D}_{GC}$, and open-system steady-state ${\cal D}_{ss}$, for two staggered fields: (a) $\delta h = 0.2$, and (b) $\delta h=0.4$. Displayed is also unperturbed value ${\cal D}_I = 0.55$. (c) Diffusion constant ${\cal D}_{ss}$ vs. $\delta h$, as obtained for sizes $L =20, 40, 60$ within the steady-state open-system approach. Shown are also values obtained via linear  $1/L \to 0$ scaling.}
\label{fig3}
\end{figure}

It is straightforward to calculate the GC averaged value of ${\cal D}_{GC}$, corresponding to $\langle m\rangle_{GC}=0$ at given size $L$. Namely, taking into account $\langle m^2 \rangle_{GC} =1/(4L)$ we get
\begin{equation}
{\cal D}_{GC}={\cal D}_0 + 4 \zeta/(L g^2)\,.
\label{dgc} 
\end{equation}
GC values are thus for weaker $g<2\sqrt{\zeta/L}$ evidently \mbox{${\cal D}_{GC} \gg {\cal D}_0$}, but should approach the latter value on increasing $L$. The results for $\delta h = 0.2, 0.4$ are presented in Fig.~\ref{fig3}. 

\section{Diffusion in open systems}

MCLM results above are based on the exact diagonalization, which restricts reachable sizes to $L \leq 32$. Larger $L$ are therefore crucial to analyse regime of weaker perturbations $g \to 0$. For this purpose, we study open Heisenberg chains, where the spin current is driven via boundary Lindblad operators, 
$L_1 = \sqrt{1+\mu} S^-_1$,
$L_2 = \sqrt{1-\mu} S^+_1$,
$L_3 = \sqrt{1-\mu} S^-_L$,
$L_4 = \sqrt{1+\mu} S^+_L$, 
simulating couplings to baths with a small difference $\mu$ in spin chemical potential \cite{znidaric11}. We use time-evolving block-decimation (TEBD) technique for vectorized density matrices \cite{zwolak04,verstraete04}, which evolves the density matrix $\rho$ under the Liouvillian $\cl{L}\rho = -i [H,\rho] + \cl{D}\rho$, $\cl{D}\rho = \sum_k L_{k}\rho L_{k}^\dagger - \frac{1}{2}\{L_{k}^\dagger L_{k}, \rho\}$, to reach the nonequilibrium steady-state $\rho_{ss}$ for small $\mu \sim 0.01$ with bond dimension $\chi_{bd}=128$. Finite bias $\mu$ yields a finite steady state spin currents $j_{ss}=\tr[\rho_{ss} j_s]$ from which diffusion constant can be obtained as ${\cal D}_{ss} = - j_{ss}/\nabla{s^z}$, where the magnetization gradient for $L \gg1$ is (nearly) homogeneous, $\nabla{s^z} = \tr[(S^z_{i+1} - S^z_i)\rho_{ss}] \sim \mu/L$. Such approach allows to deal with substantially larger $L$, but is still limited to $L \le 80$ due to very slow convergence towards the steady state as a consequence of small ${\cal D}_{ss} \ll 1$ for our parameters.

For open and closed systems, it is important to follow the $L$ dependence of results. In Fig.~\ref{fig3}(a,b) we summarize ${\cal D}$ from all methods, presented as scaling vs. $1/L$. It is evident that ${\cal D}_{ss}$ are for all cases between the canonical $m=0$ values ${\cal D}_0$ and ${\cal D}_{GC}$. The latter can be considered as an upper bound to the desired thermodynamic $1/L \to 0$ result ${\cal D}(g)$. For larger $\delta h=0.4$, ${\cal D}_{ss}$ as well as ${\cal D}_0$ and ${\cal D}_{GC}$ yield very consistent results allowing for a reliable extrapolation to $1/L \to 0$. Furthermore, for weak $\delta h=0.2$, ${\cal D}_0$ are well below ${\cal D}_{GC}$ for reachable $L$, while ${\cal D}_{ss}$ still shows relevant variation with $1/L$, even with a visible change of slope indicating on possible crossover behavior with $1/L$. 

In Fig.~\ref{fig3}(c) we present the dependence of open-system ${\cal D}_{ss}$ on $\delta h$, as calculated for different sizes, $L =20,40,60$. We notice that at weak $\delta h \to 0 $ the open-system method reproduces unperturbed value, i.e., ${\cal D}_{ss} \sim {\cal D}_I$. The continuous decrease of ${\cal D}_{ss}$ with $\delta h$ at fixed $L$ can be interpreted as (partly) persisting ballistic component due to small size $L < \lambda^*(\delta h)$. Still, the results indicate on a crossover, i.e., with increasing $L$ the decrease with $\delta h$ becomes steeper and the region with ${\cal D}_{ss} \sim {\cal D}_I$ narrower. On the other hand, ${\cal D}_{ss}$ become quite $L$ independent for larger $\delta h \ge 0.3$, as also evident from Fig.~\ref{fig3}(b). In Fig.~\ref{fig3}(c) we also present the result of a simple linear $1/L\to 0$ scaling (above the presumed crossover $\delta h \ge 0.1$ for reachable $L$), which can be regarded as an upper bound for the thermodynamic-limit value ${\cal D}_{ss}(\delta h)$. Resulting values reveal very modest variation with $\delta h$, consistent with ${\cal D}_{ss}(\delta h \to 0, L\to\infty) \ll {\cal D}_I$. 

\section{Finite wavevector and driving-field analysis} 

\begin{figure}[tb]
\includegraphics[width=0.9\columnwidth]{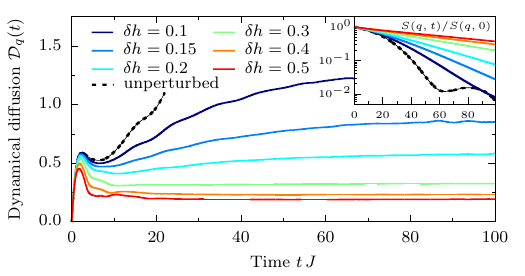}
\caption{Time-dependent diffusion ${\cal D}_q(t)$ as extracted from corresponding spin-correlation functions $S(q,t)$ (shown in log scale as inset) evaluated for $L =28$ and smallest finite $q=2\pi/L$ for staggered fields $\delta h=0-0.5$.}
\label{fig4}
\end{figure}

Transport can be also analysed by considering finite wavevector $q>0$ correlations in closed systems with PBC. In particular, we evaluate the time-dependent spin structure factor $S(q,t)$ as correlations of the operator $S^z_q = (1/\sqrt{L}) \sum_j \mathrm{exp}(i q j) S^z_j$ for the smallest nonzero $q=2 \pi/L$. The numerical procedure is again MCLM in analogy to $\tilde \sigma(\omega)$, Eq.~(\ref{kubo}), now for $S(q,\omega)$. Large number $M_L \sim 10^4$ allows to follow the time evolution for $S(q,t)$ up to $t \gg 100$. Consequently, effective (time-dependent) diffusion can be extracted as ${\cal D}_q(t) = - \dot S(q,t)/[q^2 S(q,t)]$ \cite{steinigeweg11}. It is evident that in normal systems such procedure yields constant ${\cal D}_q(t \to \infty) \sim {\cal D}$. We argue and confirm numerically that the stationary values ${\cal D}_q(t)$ can be interpreted as corresponding to the GC average for given system, since operator $S^z_q$ creates an inhomogeneous magnetization $S^z_i$ distribution, in contrast to $j_s$ in Eq.~(\ref{kubo}) related to $q=0$ and fixed $S^z_{tot}$. Results for normalized $S(q,t)$ (shown in log scale as inset) and extracted ${\cal D}_q(t)$, as calculated within MCLM for $L=28$, are presented in Fig.~\ref{fig4}. For the integrable system $\delta h =0$ it has been already noticed \cite{steinigeweg12} that ${\cal D}_q(t) \sim {\cal D}_I$ exhibit a plateau at intermediate $t<\tau_L$, corresponding to dissipationless diffusion. In contrast, for longer times, the ${\cal D}_q(t > \tau_L) \propto t$ reflects the relaxation of the Gaussian decay of $S(q,t)$ \cite{steinigeweg12}. The same behavior persists (at least for reachable $t$) for weak perturbations $\delta h$. 

On the other hand, $\delta h >0$ leads to normal diffusion manifested by the saturated values ${\cal D}_q$ at longer times $t > \tau_\lambda$ and by the exponential decay of $S(q,t)$. The saturated values ${\cal D}_q$ agrees with the GC averages corresponding to Fig.~\ref{fig2} and Eq.~(\ref{dgc}). Here, the additional comments are in order: (a) the saturation of ${\cal D}_q(t)$ is unrelated to the mechanism of anomalous integrable diffusion restricted to short $t < \tau_L$. This is evident for all $\delta h$, in particular for weak $\delta h$ where the saturation happens only at $t \gg \tau_L$. (b) Saturated ${\cal D}_q$ can be at weak $\delta h \le 0.2$ (for presented $L$) even larger than ${\cal D}_I$. (c) Large ${\cal D}_q\sim{\cal O}(1)$ are finite-size results, decreasing as $1/L$. As already discussed for the GC case, the relevant results in the macroscopic limit $L \to \infty$ are again ${\cal D}_0$, Eq.~(\ref{dgc}).

Also weak external field $F>0$ has highly nontrivial effect on an integrable chain at $\Delta >1$. In the previous study \cite{mierzejewski11}, this has been investigated within a closed system with PBC at high, but finite $T$, by introducing time-dependent flux $\phi(t)= Ft$. The spin current $j_s$ in a homogeneous canonical $m=0$ was extracted from nearly steady-state at a weak $F$, with a corresponding dc spin conductivity $\tilde \sigma_0 = T(t) j_s(t)/F$ at the effective (time-dependent) temperature $T(t)$. Results clearly indicated that $\sigma_0 < \chi_0 {\cal D}_I$, i.e., the violation of the Einstein relation in an integrable system, in particular for $F \to 0$. Here, we present additional evidence that finite $F$ breaks the integrability and consequently qualitatively changes the spin-transport response. Similarly as in Ref.~\onlinecite{mierzejewski11}, we introduce the $F>0$ via flux $\phi(t)$ (Appendix~\ref{app3}). Extracted ${\cal D}_F(t)$ exhibits qualitatively the same features as results in Fig.~\ref{fig4}. Again, saturated values ${\cal D}_F(t>\tau_L)$ at long-enough $t$ emerge due to IBP $F>0$ and are unrelated to intermediate-time $t \sim \tau_L$ saturation ${\cal D}_F(t \sim \tau_L) \sim {\cal D}_I$. Since the procedure simulates evolution of an inhomogeneous system, the stationary ${\cal D}_F$ corresponds to GC average, which can appear large, i.e.,  ${\cal D}_F >{\cal D}_I$ for weak $F$, but should decrease with $L$ to ${\cal D}_F <{\cal D}_I$ for all $F>0$. 

\section{Conclusions}
By studying the high-$T$ dynamical spin transport in perturbed anisotropic $S=1/2$ XXZ Heisenberg chain at $\Delta>1$, we establish consistent evidence that IBP lead to normal/dissipative transport, unrelated to the dissipationless diffusion ${\cal D}_I$  in the integrable system, which originates from ballistic transport in sectors with nonzero magnetization. We note that in contrast to theoretical results, the cold-atom experiments in this regime tend to support the subdiffusive transport \cite{jepsen20}, which might be due to short-time restrictions, due to presence of finite IBP, but as well due to importance of initial conditions in integrable systems. Our numerical evidence comes from the results of different approaches: canonical $m=0$ sector results ${\cal D}_0$ and GC results ${\cal D}_{GC}$ for finite closed systems; steady-state analysis in open boundary-driven systems; finite-$q$ correlations $S(q,t)$ for smallest finite $q=2\pi/L$; and response to the finite driving field $F>0$. Since this indicates the breakdown of the standard perturbation-theory approach to IBP, also the generalized Einstein relation is not valid for the integrable system. 

Several additional comments are in order: (a) Despite expected normal transport, finite but small IBP remain challenge for all numerical approaches. Namely, thermodynamic $L \to \infty$ results are expected only in systems with $L > \lambda^*(g)$. With restriction of MCLM to $L=32$ and steady-state of open systems up to $L=80$, this still represents some restriction for weakest $g \le 0.1$. (b) Clearly, the goal is to find first ${\cal D}(g)$, as extrapolated to the limit $L \to \infty$ and then ${\cal D}_0 = {\cal D}(g \to 0)$. While previous canonical $m=0$ results with finite-field driving $F>0$ \cite{mierzejewski11} seem to indicate ${\cal D}(g\to 0) \to 0$ for vanishing $F$, present MCLM analysis and even more the open-system results for homogeneous Hamiltonian IBP are plausibly consistent with a jump \cite{denardis21a}, i.e., ${\cal D}_0 >0$, but clearly ${\cal D}_0< {\cal D}_I$. (c) In this work, we focused on results for a particular anisotropy parameter $\Delta = 1.5$, but the general conclusion should remain valid for the whole regime $\Delta >1$. It is, however, expected that ${\cal D}(g)$ as well as desired ${\cal D}_0$ depend on $\Delta$ and possibly also on the actual form of IBP.

\begin{acknowledgments}
The authors thank E. Ilievski and M. \v{Z}nidari\v{c} for fruitful discussions. Our TEBD code was written in Julia \cite{bezanson17}, relying on the TensorOperations.jl. M.M. acknowledges the support by the National Science Centre, Poland via projects 2020/37/B/ST3/00020. J.H. acknowledges the support by the Polish National Agency of Academic Exchange (NAWA) under contract PPN/PPO/2018/1/00035. P.P., Z.L., and S.N. acknowledge the support by the projects N1-0088, J1-2463 and P1-0044 program of the Slovenian Research Agency.
\end{acknowledgments}

\appendix

\section{Perturbed XXZ spin chain: next-nearest-neighbor interaction} \label{app1}

\begin{figure}[!t]
\includegraphics[width=0.8\columnwidth]{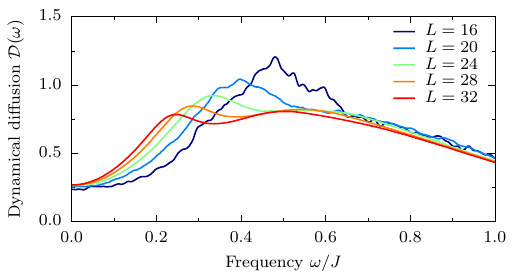}
\caption{Dynamical diffusion ${\cal D}(\omega)$ for the XXZ spin chain perturbed with NNN interaction $\Delta_2=0.4$, as obtained with MCLM for canonical $m=0$ system with different sizes $L=16-32$.}
\label{figS1}
\end{figure}

In this Appendix we discuss results for the NNN interaction $g=\Delta_2$ as IBP. The advantage of this perturbation for MCLM is that it does not break any translational symmetry, while it would require more effort to be encoded into the TEBD-based open-system analysis due to the long-range nature of exchange. It should be also recognized that at fixed $g$ such perturbation is relatively weaker to considered $g=\delta h, \delta J$ examples, i.e., it makes sense to discuss cases up to $\Delta_2 \sim 0.8$. To illustrate the finite-size effects of MCLM canonical at $m=0$, in Fig.~\ref{figS1} we present  the evolution of ${\cal D}(\omega)$ at fixed $\Delta_2 =0.4$ at various $L = 16 - 32$. While $L=16$ is still dominated by large finite-size $\omega \sim 1/L$ peak (but also noise due to small Hilbert space), with increasing $L$ the corresponding peak is shifted downwards as $1/L$ and, more importantly, loosing in intensity. At the same time there is only minor shift of the dc value ${\cal D}_0$.

\begin{figure}[!b]
\includegraphics[width=0.8\columnwidth]{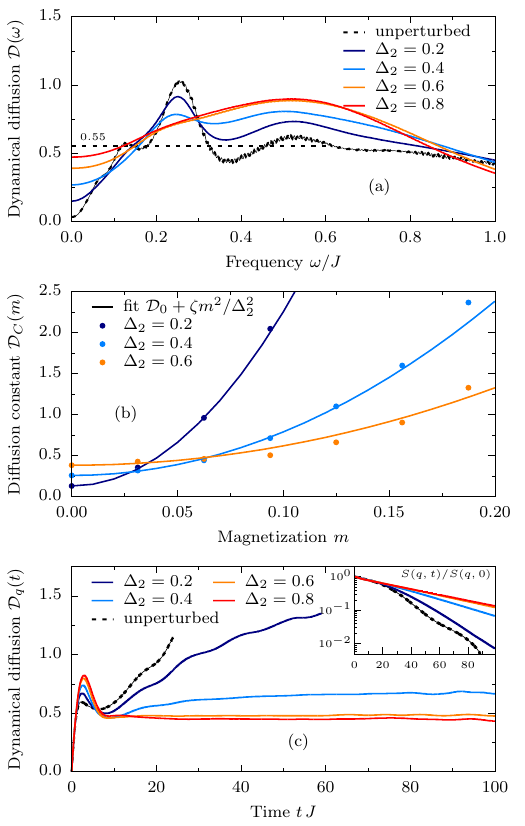}
\caption{Results for diffusion at various strengths of NNN interactions $\Delta_2=0-0.8$ for XXZ chain with $\Delta = 1.5$. (a) High-$T$ dynamical ${\cal D}(\omega)$, as calculated with MCLM on chain of $L=32$. (b) Diffusion constant ${\cal D}_C(m)$ vs. magnetization $m$. Results are fitted with parabola with fixed $\zeta = 9.0$. (c) Time-dependent ${\cal D}_q(t)$, as extracted from spin correlations $S(q,t)$ with smallest finite $q = 2\pi/L$ evaluated for $L=28$ system (inset shows the corresponding $S(q,t)$ in the log scale).}
\label{figS2}
\end{figure}

In Fig.~\ref{figS2} we summarize the NNN results in analogy to Figs.~\ref{fig1},\ref{fig2},\ref{fig4}. In Fig.~\ref{figS1}(a) we show high-$T$ canonical $m=0$ results for ${\cal D}(\omega)$ for $\Delta=1.5$ for various $\Delta_2 = 0 - 0.8$, as calculated with MCLM on $L=32$ sites. We note that smaller-$L$ results for this case were presented already in Ref.~\onlinecite{mierzejewski11}. Again, we conclude that unperturbed $\Delta_2=0$, but also weakest $\Delta_2=0.2$ result, show large finite-size $\omega \sim 1/L$ oscillations. This is not the case for larger $\Delta_2 \ge 0.4$. Evidently, for smaller $\Delta_2$ it is important that we reach as large $L$ as possible. 

In Fig.~\ref{figS2}(b) we display the canonical MCLM results for dc ${\cal D}_C(m)$ as function of magnetization $m$. Again, the variation can be captured with parabolic dependence with universal $\zeta = 9.0$, while values ${\cal D}_0$ in this case still significantly vary with $\Delta_2$. Obtained ${\cal D}_0$ are below ${\cal D}_I$ for all considered $\Delta_2 >0$. Finally, in Fig.~\ref{figS2}(c) we present the ${\cal D}_q(t)$ as extracted from spin correlation $S(q = 2\pi/L,t)$. One can notice the intermediate saturation ${\cal D}_q(t) \sim {\cal D}_I$ in unperturbed and weak $\Delta_2=0.2$ case, with saturation at longer $t$ indicating dissipative diffusion, even quantitatively consistent with the GC values following from Fig.~\ref{figS2}(b).

\section{Perturbed XXZ spin chain: staggered exchange} \label{app2}

\begin{figure}[!htb]
\includegraphics[width=0.8\columnwidth]{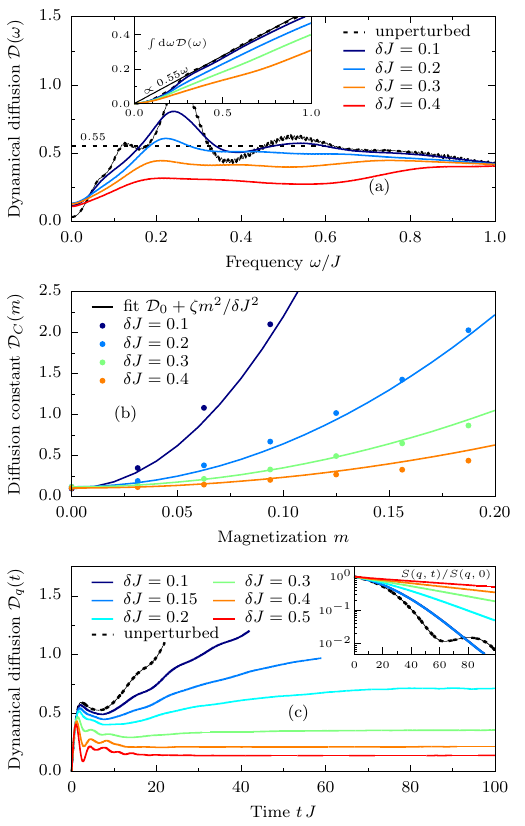}
\caption{Results for diffusion for staggered-exchange perturbation $\delta J=0-0.8$ for the XXZ chain with $\Delta = 1.5$. (a) High-$T$ dynamical ${\cal D}(\omega)$, as calculated with MCLM on chain of $L=32$. (b) Diffusion constant ${\cal D}_C(m)$ vs. magnetization $m$ with results fitted to parabola with $\zeta = 2.1$. (c) Time-dependent ${\cal D}_q(t)$ as extracted from $S(q=2\pi/L,t)$ evaluated for $L=28$ system (the inset shows corresponding $S(q,t)$ in log scale).}
\label{figS3}
\end{figure}

Results for the staggered-exchange IBP $g=\delta J$ are quite analogous and quantitatively similar to the case of staggered field $g = \delta h$. In Fig.~\ref{figS3} we display corresponding results $g=\delta J$.  Canonical $m=0$ MCLM results on $L=32$ sites for ${\cal D}(\omega)$ in Fig.~\ref{figS3}(a) for different $\delta J >0$ reveal again the evolution from finite-size dominated oscillation $\omega \propto 1/L$ at weakest $\delta J \le 0.1$, to nearly-$L$ independent spectra ${\cal D}(\omega)$ for larger $\delta J \ge 0.2$. The reduction of integrated $I(\omega)$ (in the inset) in a wide range of $\omega$ is consistent with the clear reduction of dc ${\cal D}_0 \ll {\cal D}_I$ for all $\delta J \ge 0.2$. Also similar to $\delta h$ case are conclusions for the canonical diffusion ${\cal D}_C(m)$ presented in Fig.~\ref{figS3}(b), fitted with parabola with universal $\zeta = 2.1$, as well as for the diffusion ${\cal D}_q(t)$ extracted from $S(q=2\pi/L,t)$, presented in Fig.~\ref{figS3}(c).

\begin{figure}[!htb]
\includegraphics[width=0.9\columnwidth]{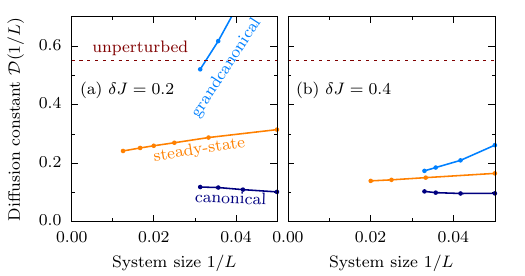}
\caption{Diffusion constant ${\cal D}$ vs. $1/L$ obtained for fixed $\Delta = 1.5$ from finite-$L$ results by three different approaches: canonical $m=0$ sector MCLM results ${\cal D}_0$, GC average over $m$ sectors ${\cal D}_{GC}$, and open-system steady-state ${\cal D}_{ss}$, for two staggered exchange: (a) $\delta J = 0.2$, and (b) $\delta J=0.4$. Displayed is also unperturbed value ${\cal D}_I = 0.55$.}
\label{figS4}
\end{figure}

Since $\delta J$ case allow for the same open-system treatment as the $\delta h$ perturbation in the main text, we summarize ${\cal D}$ calculated by three different methods in Fig.~\ref{figS4} for two strengths $\delta J = 0.2, 0.4$, in analogy to Fig.~\ref{fig3} for $\delta h$ perturbation. While for stronger $\delta J=0.4$ all three methods yield quite converged results, for $\delta J =0.2$ MCLM ${\cal D}_{GC} \gg {\cal D}_0$ are still quite apart for reachable $L$, whereas steady-state ${\cal D}_{ss}$ is for largest $L=80$ apparently much closer to convergence. 

\section{Diffusion in the presence of finite driving field} \label{app3}

\begin{figure}[!htb]
\includegraphics[width=0.8\columnwidth]{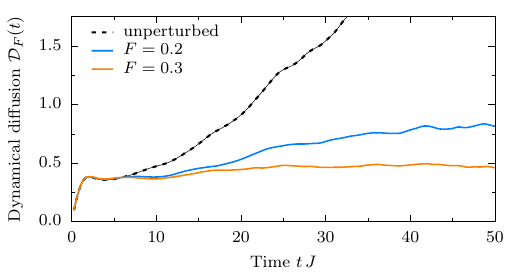}
\caption{Time-dependent diffusion ${\cal D}_F(t)$ as extracted from the time evolution of the spin-density profile with the smallest finite $q=2\pi/L$ for the system of $L=28$ sites with $\Delta =1.5$ for various driving fields $F=0,0.2,0.3$.} 
\label{figS5}
\end{figure}
The conductivity is defined as a ratio of the steady current $j_s$ and the steady driving field $F$ for vanishingly small driving, dc $\sigma_0=\lim_{F \to 0} j_s/F$. As a consequence the system's response should be determined in the presence of small but nonzero driving. In generic nonintegrable systems, the current-current correlation functions do not exhibit any discontinuity upon introducing weak perturbations, so that dc $\sigma_0$ can be calculated strictly for $F=0$. However, the correlation functions in the easy-axis XXZ model may change discontinuously upon introducing a perturbation, whereby the role of 
finite driving $F$ might be different from Hamiltonian IBP discussed in the main text. To consider $F>0$  we study the XXZ model with time dependent Hamiltonian 
\begin{eqnarray}
H(t\le 0 ) & =& J \sum_{i} \bigl[ \frac{1}{2} ( S^+_{i+1} S^-_i + \mathrm{H.c.} ) + \Delta S^z_{i+1} S^z_i \bigr] \nonumber \\
&& + \sum_i \frac{1}{2} \cos \left( q i \right) S^z_i, \label{sh0} \\
H(t>0)&=& J \sum_{i} \bigl[ \frac{1}{2} (e^{iFt} S^+_{i+1} S^-_i + \mathrm{H.c.} ) + 
\Delta S^z_{i+1} S^z_i \bigr]. \label{sh1} \nonumber \\
\end{eqnarray}
Using MCLM, we generate the initial microcanonical state at time $t=0$, $| \psi_0 \rangle$, corresponding to high, but finite temperature, $T \simeq 5$. The last term in Eq.~(\ref{sh0}) is responsible for the spin-density wave, $\langle S^z_i \rangle_0=\langle \psi_0 | S^z_i | \psi_0 \rangle \propto \cos(q i)$, where we choose the smallest finite $q=2\pi/L$. Next, at $t>0$ this term is switched off while the driving field $F>0$ is switched on, and $|\psi_t \rangle $ evolves under the Hamiltonian (\ref{sh1}). During time propagation we calculate the variance of the spatial spin profile, $(\delta S_t)^2 =(1/L) \sum_i \langle \psi_t | S^z_i | \psi_t \rangle^2$. For normal diffusive systems without driving the amplitude of the spin-density wave decays exponentially in time, $\delta S_t=\delta S_0 \exp(-{\cal D}_F q^2 t)$ hence the effective time-dependent diffusion constant can be obtained as \mbox{${\cal D}_F(t) = - \delta \dot S_t /[q^2 \delta S_t ]$}. It is evident that the present setup which is based on time propagation of the spin-density profile, is equivalent to the analysis based on the spin structure factor $S(q,t)$ of the main text. The disadvantage of the present approach is that the dynamics is evaluated at finite temperature ($T \gg 1 $) and results may be affected by nonlinear contributions which are beyond the linear-response theory. However, the time-propagation method allows for time-dependent Hamiltonians and one may include the steady driving as in the Hamiltonian (\ref{sh1}). The resulting ${\cal D}_F(t)$ is shown in Fig.~\ref{figS4} for the integrable case $F=0$, as well as for driven systems with $F=0.2$ and $F=0.3$. It is clear that nonzero driving affects the spin dynamics in the same way, as IBP shown in Fig.\ref{fig4} as well as in Figs.~\ref{figS2}(c) and Fig.~\ref{figS3}(d). 
 
\bibliography{manuandiff}

\end{document}